\documentclass[aps,prb,twocolumn,superscriptaddress,floatfix]{revtex4}
\usepackage{graphicx}
\begin{document}

\title{Degenerate ground states and nonunique potentials:
breakdown and restoration of density functionals}

\author{K. Capelle}
\affiliation{Departamento de F\'{\i}sica e Inform\'atica,
Instituto de F\'{\i}sica de S\~ao Carlos,
Universidade de S\~ao Paulo,
Caixa Postal 369, 13560-970 S\~ao Carlos, SP, Brazil}

\author{C. A. Ullrich}
\author{G. Vignale}
\affiliation{Department of Physics and Astronomy,
University of Missouri-Columbia, Columbia, Missouri 65211, USA}

\date{\today}

\begin{abstract}
The Hohenberg-Kohn (HK) theorem is one of the most fundamental theorems of
quantum mechanics, and constitutes the basis for the very successful
density-functional approach to inhomogeneous interacting many-particle systems.
Here we show that in formulations of density-functional theory (DFT) that
employ more than one density variable, applied to systems with a degenerate
ground state, there is a subtle loophole in the HK theorem, as all mappings
between densities, wave functions and potentials can break down. Two weaker
theorems which we prove here, the {\em joint-degeneracy theorem} and the
{\em internal-energy theorem}, restore the internal, total and exchange-correlation energy
functionals to the extent needed in applications of DFT to atomic, molecular
and solid-state physics and quantum chemistry. The joint-degeneracy theorem
constrains the nature of possible degeneracies in general many-body systems.
\end{abstract}


\maketitle

\newcommand{\be}{\begin{equation}}
\newcommand{\ee}{\end{equation}}
\newcommand{\bea}{\begin{eqnarray}}
\newcommand{\eea}{\end{eqnarray}}
\newcommand{\bi}{\bibitem}

\renewcommand{\r}{({\bf r})}
\newcommand{\rp}{({\bf r'})}

\newcommand{\ua}{\uparrow}
\newcommand{\da}{\downarrow}
\newcommand{\la}{\langle}
\newcommand{\ra}{\rangle}
\newcommand{\dg}{\dagger}

{\em Introduction.}
Quantum mechanics is based on the assumption that all information that
one can, in principle, extract from a system in a pure state at zero
temperature is contained in its wave function. In nonrelativistic quantum
mechanics the wave function obeys Schr\"odinger's equation,\cite{schroed}
which implies a powerful variational principle according
to which the ground-state wave function minimizes the
expectation value of the Hamiltonian. This variational principle was used
by Hohenberg and Kohn (HK)\cite{hk} to show that the entire
information contained in the wave function is also contained in the
system's ground-state particle density $n\r$.

HK established the existence of two mappings,
\be
v\r \stackrel{1}{\Longleftrightarrow} \Psi({\bf r}_1,\ldots {\bf r}_N)
\stackrel{2}{\Longleftrightarrow} n\r,
\label{maps}
\ee
where the first guarantees that the single-particle potential is a unique
functional of the wave function, $v[\Psi]$, and the second implies that the
ground-state wave function is a unique functional of the ground-state density,
$\Psi[n]$. Taken together, both mappings are encapsulated in the single
statement that the single-particle potential is a unique density functional
$v[n]$. In this formulation, the HK theorem forms the basis of the
spectacularly successful approach to many-body physics, electronic-structure
theory and quantum chemistry that became known as density-functional theory
(DFT).\cite{kohnrmp,dftbook,parryang}

Mapping 2 was originally proven by contradiction\cite{hk} and later
by constrained search.\cite{levylieb} Note that, in spite of occasional
statements to the contrary in the literature, neither proof directly
proves the combined mapping, and thus the existence of the functional $v[n]$.
This requires additionally mapping 1, which in the case of density-only DFT
is proven by inverting Schr\"odinger's equation\cite{dftbook,levyaqc}
\be
\hat{V} =\sum_i v({\bf r}_i) 
= E_k - \frac{(\hat{T}+\hat{U})\Psi_k({\bf r}_1,\ldots {\bf r}_N)}
{\Psi_k({\bf r}_1,\ldots {\bf r}_N)}.
\label{invert}
\ee
According to the preceding equation, any eigenstate $\Psi_k$
determines the single-particle potential $v\r$ up to an additive constant,
the corresponding eigenenergy.\cite{footnote1}
The ground-state version of this equation is all one needs to
prove mapping 1, and thus of the combined mapping
$n\r \Rightarrow v\r+const$, which we write as $v=v[n]$. The constant
can always be absorbed in the definition of the zero of energy, but the
resulting nonuniqueness of the mapping $v[n]$ gives rise to
derivative discontinuities\cite{discont} that crucially contribute to
observables such as the band gap, chemical hardness, electron affinities,
electron transfer energies, among others.

The two classic proofs of the HK theorem, by contradiction
and by constrained search, have been extended to a wide variety of systems,
including spin magnetization (in spin-density-functional theory --
SDFT \cite{vbh,gl}) and orbital currents (in current-density-functional
theory -- CDFT\cite{vr}), among others.
Further scrutiny, however, has led to the discovery of situations in which
each of the two mappings breaks down.

{\em Breakdown of mappings: Degeneracy.}
Mapping 2, $\Psi[n]$, breaks down in the presence of
degenerate ground states, where neither the constrained search proof
nor the proof by contradiction provide a unique wave function among the
degenerate manifold, because the strict inequality of the underlying
variational principle is replaced by the weaker relation
less-or-equal. In the case of the proof by contradiction, the
contradiction simply does not follow unless one has a strict inequality,
and in the proof by constrained search there is no guarantee that the
search delivers only one wave function for a specified density.

\begin{figure}
\centering
\includegraphics[height=50mm,width=70mm,angle=0]{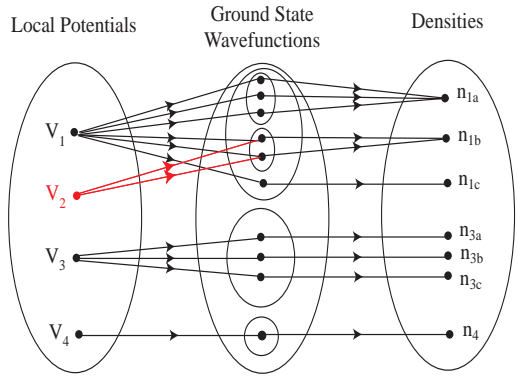}
\caption {\label{fig1} (color online)
Schematic illustration of the breakdown of mappings that occurs in the
presence of degeneracy (black) and the additional complication posed
by nonuniqueness (red). $V$ and $n$ stand generically for sets of conjugate
potentials and densities (e.g., $v_\ua,v_\da$ and $n_\ua,n_\da$ in SDFT).
Large ovals are sets of functions, medium-size ovals collect degenerate
wave functions, and small ovals enclose degenerate wave functions that give
rise to the same density.}
\end{figure}

What remains from the proofs is that the ground-state density uniquely
determines a manifold of degenerate states, $\{\Psi_i\}[n]$, but not all of
these states individually.\cite{footnote2} Any member of this manifold,
however, still uniquely determines the potential, since any of them can be
used in Eq.~(\ref{invert}). Hence, even in the presence of degeneracy, the
mapping $v[\Psi_i]$, and thus $v[n]$, still exists. This
situation is illustrated in black in Fig. \ref{fig1}.


To proceed from establishing mappings to a practical density-functional
theory, one must define the total-energy functional $E_v[n]$, the universal
internal-energy functional $F[n]$, and the exchange-correlation energy
functional $E_{xc}[n]$. The Kohn-Sham formulation additionally
requires the noninteracting kinetic-energy functional $T_s[n]$. 21 years after
proving the original HK theorem, Kohn\cite{kohnhighlight} showed how these
functionals can be defined even if degeneracy renders the original proof
ineffective.\cite{dftbook} Since all degenerate wave functions by definition
yield the same ground-state energy $E$, one can directly define the functional
\be
F[n]:= E - \int d^3r\, n\r v[n]\r = E - V[n].
\ee
Conventionally, this functional is defined as $F[n] = T[n]+U[n]$, but the
information that the kinetic energy $T$ and interaction energy $U$ are
density functionals is only available if the second mapping, $\Psi[n]$,
holds, and cannot be taken for granted in the presence of degeneracy.
By contrast, the alternative definition above only requires the mapping
$v[n]$ to establish the existence of the universal internal-energy functional
$F[n]$. Similarly, in the noninteracting KS system one defines
\be
T_s[n]:=E_s-\int d^3r\, n\r v_s[n]\r = E_s - V_s[n]
\ee
where $v_s\r$ is the effective KS potential and $E_s$ the
ground-state energy of the KS system, i.e., the sum of the KS eigenvalues.
The $F$ and $T_s$ functionals defined in this way can be used to
establish a density variational principle for $E_v[n]$
and to define the exchange-correlation energy $E_{xc}[n]=F[n]-T_s[n]-E_H[n]$,
where $E_H$ is the Hartree energy.\cite{dftbook}

Thus tamed, degeneracy actually becomes helpful in further
strengthening the foundations of DFT: on a lattice, any density can be written
as a linear combination of densities arising from ensembles of degenerate
ground states of a local potential, thus solving the discretized
$v-$representability problem.\cite{ullrichkohn}

{\em Breakdown of mappings: Nonuniqueness.}
It is known\cite{harriman} at least since 1983 that mapping 1,
$v[\Psi]$, breaks down if finite basis sets are used to represent the wave
functions. Harriman\cite{harriman} gives both general arguments and an
explicit example illustrating this breakdown.
This breakdown of mapping 1 is the only one occuring already in the
charge-density-only formulation of DFT, and it is manifestly an
artifact of the use of a finite basis set.

In multi-density DFTs, such as SDFT and CDFT, the mapping between
the set of effective potentials and the set of ground-state densities
can break down even in the complete basis-set limit, because inversion of
Schr\"odinger's equation
does not establish a unique relation between the set of densities and the
set of conjugate potentials. This is the so-called nonuniqueness
problem of SDFT (and CDFT and others). Following an early observation of
the problem by von Barth and Hedin,\cite{vbh} the problem has been shown
to be fundamental and pervasive in recent work by Eschrig and Pickett\cite{ep}
and by two of us,\cite{nonunprl} who provided explicit examples of
different SDFT potentials sharing the same ground-state wave function.
Ref.~\onlinecite{nonunprl} proposed a classification of nonuniqueness into
systematic (arising from the existence of certain constants of motion)
and accidental (arising from special features of the ground state). In both
cases, the nonuniqueness is associated with the external potential.
Since the mapping $\Psi[n]$ remains intact, and internal-energy functionals
can be defined exclusively in terms of wave functions,
\be
F[n]= \la \Psi[n]| \hat{T}+\hat{U}| \Psi[n]\ra ,
\label{excdef}
\ee
the functionals $E_v[n]=F[n]+V[n]$, $T_s[n]=\la\Phi[n]|\hat{T}|\Phi[n]\ra$,
and $E_{xc}[n]= F[n]-E_H[n]-T_s[n]$ still exist.

Additional examples of both systematic and accidental nonuniqueness were
found in CDFT\cite{nonunprb} and in SDFT on a lattice.\cite{nonuncau}
The extent to which nonuniqueness of the potentials affects various types of
applications of multi-density DFTs, as well as possible remedies, are discussed
in Refs.~\onlinecite{nonunprb,nonuncau,nonunrefs}.

{\em Breakdown of mappings: Nonuniqueness {\em and} degeneracy.}
We have just seen that in the presence of nonuniqueness the mapping $v[\Psi]$
breaks down, whereas in the presence of degeneracy the mapping $\Psi[n]$ breaks
down. Interestingly, a crucial fact has been overlooked in the standard
analysis of either degeneracy or nonuniqueness: These complications can occur
simultaneously!
If a system with a degenerate ground state is treated with SDFT or any
other formulation of DFT suffering from a nonuniqueness problem, none of
the mappings holds, and no conventional HK theorem exists. In fact, it
is $\Psi[n]$ that was used above to define $F[n]$, $T_s[n]$ and $E_{xc}[n]$
in the absence of $v[\Psi]$ (nonuniqueness), while $v[\Psi]$ guaranteed the
existence of $v[n]$ and thus of $F[n]$, $T_s[n]$ and $E_{xc}[n]$ in the
absence of $\Psi[n]$ (degeneracy). If both $\Psi[n]$ and $v[\Psi]$ break
down, it seems nothing is left. The breakdown of both mappings is illustrated
in Fig. \ref{fig1}. Three simple examples are given below.

Our first example is an extension of the case of the noninteracting Li
atom\cite{footnote2} to collinear SDFT. If the spin degree of freedom is
included, each of the four degenerate states\cite{footnote2} is additionally
twofold degenerate with respect to $S_z$. The set of external potentials
$B=0,v=3/r$ thus has an 8-fold degenerate ground-state manifold. The
Slater determinants formed from the configurations $1s^2 2p^+\ua$ and
$1s^2 2p^-\ua$ have the same charge and spin densities. Again,
we see that in the presence of a degenerate ground state the densities do not
uniquely determine the wave function.
Differently from above, in SDFT we can now also consider the alternative set
of external potentials $B'=const\neq 0,v=3/r$. The spin-only magnetic field
$B$ simply splits the ground-state manifold into two, one comprising the four
spin-up configurations, the other the four spin-down configurations. The new
ground state will be in the spin-up manifold, where the configurations
$1s^2 2p^+\ua$ and $1s^2 2p^-\ua$ remain and still yield the same densities.
From the point of view of the mapping between densities and potentials, this
is simply the by now well known\cite{ep,nonunprl,nonunprb,nonuncau,nonunrefs}
nonuniqueness
of the potentials of SDFT with respect to a weak collinear magnetic field.
The full situation, however, is now one in which the densities do not
determine the wave functions but only a (ground-state) manifold of them, and
some members of these manifolds are ground states in more than one set
of external potentials. The functionals $\Psi[n]$, $V[\Psi]$ and $V[n]$ thus
do not exist.

Consider next an interacting atom in an $S=1$, $L=1$ state. Concrete
examples are $\rm ^6C$ and $\rm ^{14}Si$ (with term $^3P_0$) and
$\rm ^8O$ and $\rm ^{16}S$ (with term $^3P_2$). In the set of
external potentials $B=0,v=Z/r$ the ground state of such systems is
$(2L+1)(2S+1)=9$-fold degenerate. Let's denote the members of this
manifold as $\Psi_{L_z,S_z}$. Several of these, such as
$\Psi_{1,1}$ and $\Psi_{-1,1}$ have the same charge and spin densities.
Hence, we have another situation in which these densities do not
determine the wave functions but only the manifold.
Now consider the same system in external potentials $B'=const\neq 0$ and
$v=Z/r$. The states $\Psi_{1,1}$, $\Psi_{0,1}$ and $\Psi_{-1,1}$ remain
degenerate ground states in this new set of potentials, and the density and
spin density of the first and the last are still the same as for $B'=0$.
Hence, as in other examples of nonuniqueness, knowledge of this state alone
does not determine the external potentials.
Upon combining both observations we find that to a given set of ground-state
densities $(n,m)$ there may correspond more than one degenerate wave function
(all in external potentials $B=0,v=Z/r$), and all of these wave functions are
also degenerate ground-states of the different set of external potentials
$(B',v)$. Again, the functionals $\Psi[n]$, $V[\Psi]$ and $V[n]$ do not exist.

Lastly, we discuss a modification of the one-electron example by von Barth
and Hedin.\cite{vbh} Consider a single electron in the presence of an external
4-potential $w_{\alpha \beta}({\bf r})
= V({\bf r})\delta_{\alpha\beta} - [{\bf B}({\bf r}) \cdot \vec{\sigma}]_{\alpha\beta}$,
where $\vec{\sigma}$ is the vector of Pauli matrices. Let $w_{\alpha \beta}({\bf r})$
be uniform along one spatial direction (say, $x$), with periodic boundary conditions along
that direction separated by a distance $L$  (which is topologically equivalent to confining
the electron on a ring). The  two-fold degenerate ground state of the Hamiltonian
$H_{\alpha \beta} = -\frac{\hbar^2}{2m} \delta_{\alpha \beta} + w_{\alpha \beta}({\bf r})$
is given by $\Psi^{\pm}_\alpha({\bf r}) = e^{\pm i k x}\psi_\alpha(y,z)$, $k=2\pi/L$,
with both ground states producing the same density. Furthermore,
both ground states $\Psi^{\pm}_\alpha({\bf r})$ are invariant under perturbations
$w'_{\alpha \beta}({\bf r})
= V'({\bf r})\{\delta_{\alpha\beta} -  [{\bf m}({\bf r})\cdot \vec{\sigma}/n({\bf r})]_{\alpha\beta}\}$,
where $n({\bf r})$ and ${\bf m}({\bf r})$ are the ground-state density and magnetization,
and $V'({\bf r})$ is an arbitrary (but not too large) scalar potential function.
Thus, we have found another case where both mappings (\ref{maps}) break down.

\begin{figure}
\centering
\includegraphics[height=40mm,width=80mm,angle=0]{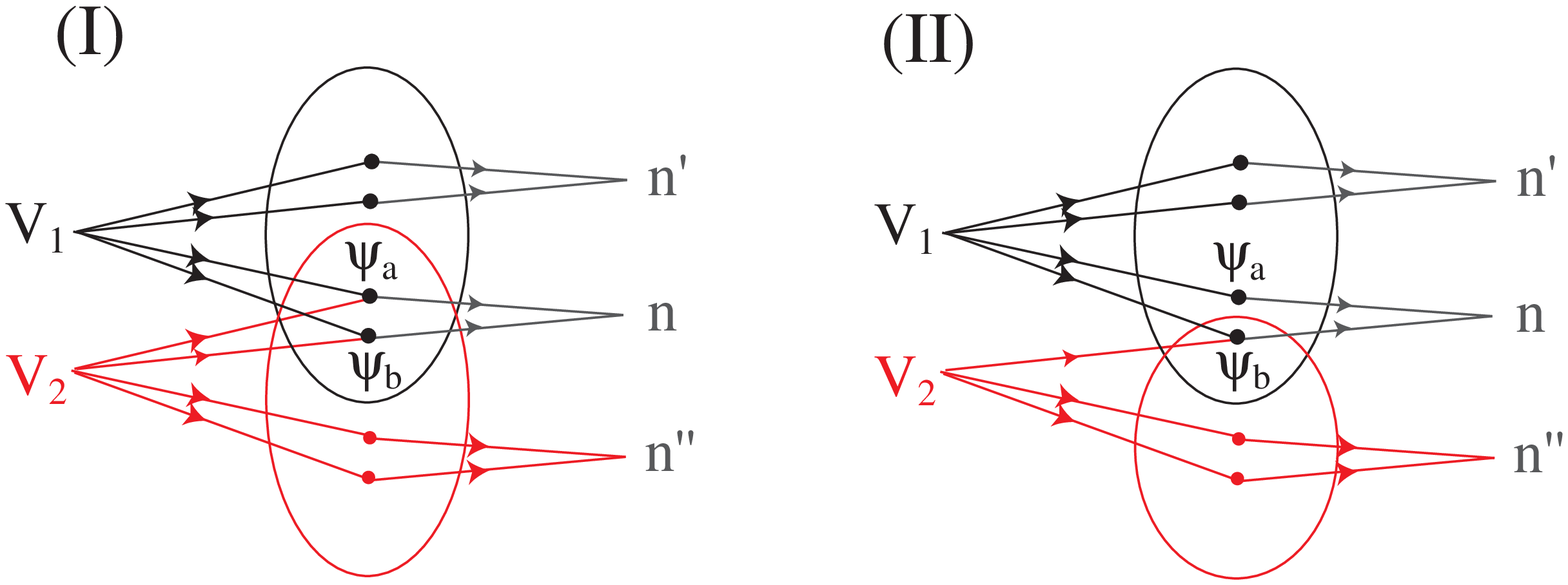}
\caption {\label{fig2} (color online)
The joint-degeneracy theorem (case I) and a situation excluded by it
(case II). Colored ovals enclose degenerate wave functions coming from
a local potential.}
\end{figure}

{\em Restoration of energy functionals.}
The question then arises whether the energy functionals $E_v[n]$, $F[n]$,
$T_s[n]$ and $E_{xc}[n]$ can still be defined, even in the absence of all
mappings that are conventionally considered the content of the HK theorem.
To answer affirmatively, we consider two distinct cases, represented in
Fig.~\ref{fig2}. In case I, $\Psi_a$ and $\Psi_b$, which both produce density
$n$, are degenerate ground states in potentials $V_1$ and $V_2$. In
case II only $\Psi_b$ is common ground state of both potentials, whereas
$\Psi_a$ is ground state only of $V_1$, but either an excited state or not
even an eigenstate at all in $V_2$.
We first prove that case II cannot occur. In potential $V_1$, we define the
internal energy (not yet a functional of any density) as
\be
F_1=E_1-\int d^3 r v_1\r n\r .
\label{f1def}
\ee
In potential $V_1$, $\Psi_a$ and $\Psi_b$ are degenerate, so that
$E_1=\la \Psi_a | \hat{T}+\hat{U} +\hat{V}_1|\Psi_a\ra
=\la \Psi_b | \hat{T}+\hat{U} +\hat{V}_1|\Psi_b\ra$. Since both also produce
the same density, the expectation value of $\hat{V}_1$ with $\Psi_a$ and
$\Psi_b$ is the same, so that
\be
\la \Psi_a | \hat{T}+\hat{U}|\Psi_a\ra
= \la \Psi_b | \hat{T}+\hat{U}|\Psi_b\ra = F_1
\label{f1tu}
\ee
is independent of the choice of wave function. Next,
applying the variational principle to the Hamiltonian of system 2, we have
\be
\la \Psi_a | \hat{T}+\hat{U} +\hat{V}_2|\Psi_a\ra > E_2,
\ee
with a strict inequality, as by assumption $\Psi_a$ in potential 2 is not
degenerate with the ground state of that system. From Eq.~(\ref{f1tu}),
and again making use of the fact that $\Psi_a$ and $\Psi_b$ produce the
same density, we can write this as
\be
\la \Psi_b | \hat{T}+\hat{U} +\hat{V}_2|\Psi_b\ra = E_2 > E_2.
\ee
The contradiction proves that case II cannot occur. This result is completely
general, and implies the following theorem:
{\em Consider two degenerate ground-state wave functions in potential $V_1$,
$\Psi_a$ and $\Psi_b$. The constraint that these two wave functions have the
same density guarantees that in any other potential $V_2$ either both are
ground states (and thus also degenerate) or none of them is.} We call this
the {\em joint-degeneracy} theorem. Note that all of our explicit examples
above respect the joint-degeneracy theorem.

Even though one could formally define an $F[n]$ functional in case II, we have
just shown that that case cannot occur, so we only need to establish the
existence of $F[n]$ in case I. In analogy to Eq.~(\ref{f1def}),
we define in potential $V_2$
\be
F_2=E_2-\int d^3 r v_2\r n\r.
\label{f2def}
\ee
Since $\Psi_a$ and $\Psi_b$ are degenerate also in potential $V_2$,
we have $E_2=\la \Psi_a | \hat{T}+\hat{U} +\hat{V}_2|\Psi_a\ra
=\la \Psi_b | \hat{T}+\hat{U} +\hat{V}_2|\Psi_b\ra$, resulting in
\be
F_2 = \la \Psi_a | \hat{T}+\hat{U}|\Psi_a\ra
= \la \Psi_b | \hat{T}+\hat{U}|\Psi_b\ra,
\label{f2tu}
\ee
where we again used that $\Psi_a$ and $\Psi_b$ yield the same density.
This is the same equation obtained above for $F_1$. Again, this result is
completely general, implying the following theorem:
{\em Regardless of any possible degeneracy or nonuniqueness, two systems with
same ground-state density have the same internal energy $F$. Hence, the
functional $F[n]$ exists and is universal, i.e., independent of the
potentials.} This {\em internal-energy theorem} is consistent with the
constrained-search formulation of DFT,\cite{levylieb} which defines
$F[n]:=\min_{\Psi\to n} \la \Psi | \hat{T}+\hat{U} | \Psi \ra$, although in
the presence of degeneracy this definition cannot be used to define $\Psi[n]$.

Since the noninteracting kinetic energy $T_s$ is the internal energy of
the Kohn-Sham system, it is also a well-defined density functional, and
$E_{xc}[n]=F[n]-E_H[n]-T_s[n]$ can be constructed as usual. Finally, for a
given external potential, the functional $E_v[n]$ then obviously also exists.


{\em Conclusions.}
We have shown both by general arguments and by specific examples
that in the case of degeneracy in multi-density DFTs all three mappings,
$\Psi[n]$, $V[\psi]$ and $V[n]$, and thus the entire body of
information usually considered the content of the HK theorem, break down.
The weaker {\em joint-degeneracy } and {\em internal-energy theorems},
however, still allow the definition of the internal-energy functional $F[n]$,
and thus also the functionals $T_s[n]$, $E_v[n]$ and $E_{xc}[n]$. Practical
DFT, which assumes the existence of these functionals, is thus largely
unaffected by the breakdown of the various mappings. However, we stress that
we have only proven existence of the functionals, not their
differentiability. In fact, in the presence of nonuniqueness, all these
functionals are expected to display derivative discontinuities.

K.C. was supported by FAPESP and CNPq. C.A.U. acknowledges support by
DOE Grant No. DE-FG02-05ER46213,
NSF Grant No. DMR-0553485 and Research Corporation.
GV acknowledges  support from NSF Grant No. DMR-0313681.

\newpage

\newpage
\hspace*{1cm}
\newpage

\end{document}